\documentclass[5p,twocolumn,times,number]{elsarticle}
\usepackage{amsmath}
\usepackage{amssymb}
\usepackage{graphics}
\usepackage{graphicx}
\usepackage{epsfig}
\usepackage{dcolumn}
\usepackage{bm}
\usepackage{lineno}

\begin{document}
\begin{frontmatter}

\title{Study of the characteristics of GEM detectors for the future FAIR experiment CBM}
\author[label1]{S.~Biswas\corref{cor}}
\ead{S.Biswas@gsi.de}
\author[label1,label2]{A.~Abuhoza}
\author[label1]{U.~Frankenfeld}
\author[label1]{J.~Hehner}
\author[label1]{C. J.~Schmidt}
\author[label3]{H.R.~Schmidt}
\author[label1]{M.~Tr\"{a}ger}
\author[label4]{S.~Colafranceschi}
\author[label4]{A.~Marinov}
\author[label4]{A.~Sharma}

\cortext[cor]{Corresponding author}

\address[label1]{GSI Helmholtzzentrum f\"{u}r Schwerionenforschung GmbH, Darmstadt, Germany-64291}
\address[label2]{King Abdulaziz City for Science and Technology (KACST),Riyadh, Saudi Arabia}
\address[label3]{Eberhard-Karls-Universit\"{a}t, T\"{u}bingen, Germany}
\address[label4]{Physics Department, CERN - Geneva, Switzerland}

\begin{abstract}
Characteristics of triple GEM detector has been studied systematically. The variation of the effective gain and energy resolution of GEM with variation of the applied voltage has been measured with Fe$^{55}$ X-ray source for different gas mixtures and with different gas flow rates. Long-term test of the GEM has also been performed.   
\end{abstract}
\end{frontmatter}

\section{Introduction}
\label{}
The Compressed Baryonic Matter (CBM) experiment at the future Facility for Antiproton and Ion Research (FAIR) in Darmstadt, Germany will use proton and heavy ion beams to study matter at extreme conditions \cite{CBM,FAIR}. The CBM experiment at FAIR is designed to explore the QCD phase diagram in the region of high baryon densities \cite{CBM2008}. This will only be possible with the application of advanced instrumentation, including highly segmented and fast gaseous detectors. 

Gas Electron multipliers (GEM) will be used in CBM Muon Chamber (MUCH) located downstream of the Silicon Tracking System (STS) of the CBM experiment along with other sophisticated detectors \cite{FS97}. The main goal of MUCH is to detect the dimuon signals arising from the decay of the low mass vector mesons and those from the decay of charmonia produced in the heavy ion collisions at FAIR.

In GSI detector laboratory an R\&D effort is launched to study the characteristics of GEM detectors for the CBM experiment. The primary goals of this R\&D program are: (a) to verify the stability and integrity of the GEM detectors over a period of time, during which a charge density of the order of several Coulomb/cm$^{2}$ is accumulated in the detector; (b) to establish the functioning of triple GEM as a precise tracking detector under the extreme condition of the CBM experiment; (c) to study usual parameters e.g., efficiency, rate capability, long term stability, spark probability by varying conditions like temperature, gas composition or radiation dose.

In this article, we would like to present the initial results of characterisation of triple GEM detectors for CBM.

\section{Detector descriptions and tests}
\label{}
\begin{figure}[h!]
\centering
\includegraphics[scale=0.4]{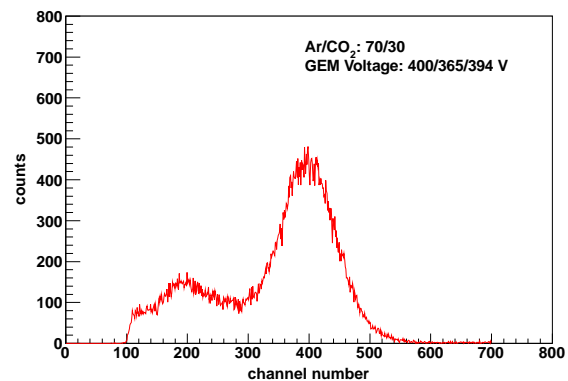}\\
\caption{Fe$^{55}$ X-ray spectrum obtained with triple GEM detector.} \label{fig:6}
\end{figure}

A triple GEM detector, consisting of double mask foils, obtained from CERN has been studied systematically. The drift, transfer and induction gaps of the detector are 3 mm, 2 mm and 2 mm respectively. The voltage to the drift plane and individual GEM plates has been applied through a voltage divider chain. Although there is a segmented readout pad the signal in this study was obtained from all the pads summed by a add up board and a single input is fed to a charge sensitive preamplifier. After that a PXI LabVIEW based data acquisition system is used. The gain of the detector has been measured by obtaining the mean position of 5.9 keV peak of Fe$^{55}$ X-ray spectrum with Gaussian fitting. A typical spectrum recorded with a Fe$^{55}$ source is shown in Fig.~\ref{fig:6}. The variation of the gain and energy resolution of this detector with that of the applied voltage has been measured for different gas mixtures and with different gas flow rates.
\section{Results}
\label{}
\begin{figure}[h!]
\centering
\includegraphics[scale=0.32]{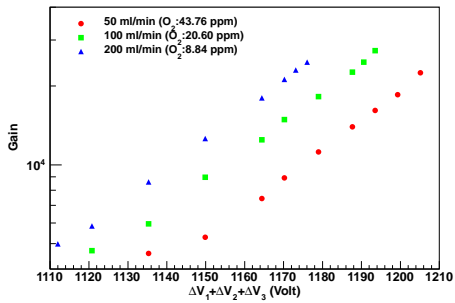}\\
\caption{Gain vs. global voltage. } \label{fig:1}
\end{figure}

\begin{figure}[h!]
\centering
\includegraphics[scale=0.32]{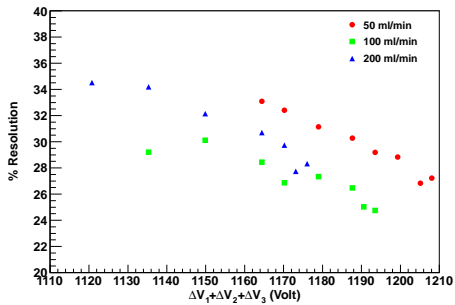}\\
\caption{Resolution vs. global voltage.} \label{fig:2}
\end{figure}

The detector has been operated with Argon and CO$_{2}$ in 60/40, 70/30, 80/20 mixing ratios with different flow rates e.g. 50~ml/min, 100~ml/min and 200~ml/min. The variation of effective gain and energy resolution with that of global voltage ($\Delta$V$_{1}$+$\Delta$V$_{2}$+$\Delta$V$_{3}$) are shown in Fig.~\ref{fig:1} and Fig.~\ref{fig:2} respectively for the gas mixture Ar/CO$_{2}$ in 70/30 with different flow rates. It is clear that the gain increases exponentially and the energy resolution improves with the increase of voltage. The gain is relatively higher with the higher gas flow rate. This is due to the less oxygen contents (shown in Fig.~\ref{fig:1}) during higher flow rate. Same behavior is also observed for the other two gas mixtures.

\begin{figure}[h!]
\centering
\includegraphics[scale=0.32]{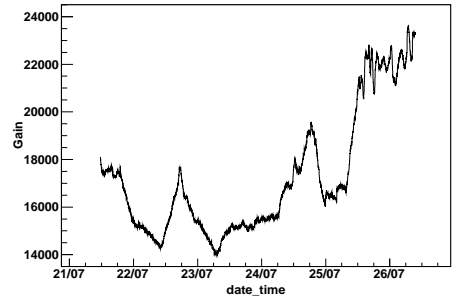}\\
\caption{Gain as a funcion of period of operation. } \label{fig:3}
\end{figure}

\begin{figure}[h!]
\centering
\includegraphics[scale=0.32]{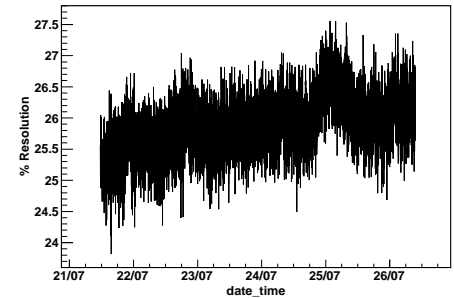}\\
\caption{Resolution as a funcion of period of operation.} \label{fig:4}
\end{figure}

The detector is operated continuously for five days with Ar/CO$_{2}$ in 70/30 ratio with 100 ml/min flow rate at a global voltage 1176 V. The gain and resolution as function of time are shown in Fig.~\ref{fig:3} and Fig.~\ref{fig:4} respectively. 

\begin{figure}[h!]
\centering
\includegraphics[scale=0.18]{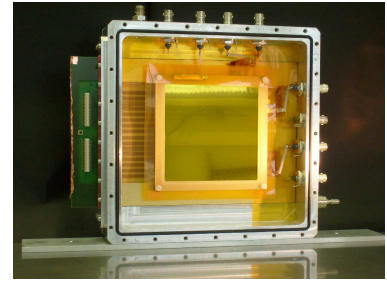}\\
\caption{New GEM box designed at GSI.} \label{fig:5}
\end{figure}
\section{Conclusions and outlook}
\label{}
Variation of gain and resolution with voltage for different gas mixtures and with different gas flow rates has been measured in GEM detectors. It is clear that oxygen content in the gas has a very important role in GEM performance even in a few ppm level.

In an initial continuous operation of five days the gain fluctuates with time, and resolution varies about 4\%. Further investigation is necessary to correlate this variation with other ambient parameters.  
 
Fabrication of new 10~cm~$\times$~10~cm GEM detector with newly designed box (shown in Fig.~\ref{fig:5})  and their characterisation is also carried out. Further studies are still in progress and will be communicated at a later stage.

\section{Acknowledgement}
\label{}
We are thankful to Dr. Ingo Fr\"{o}hlich of University of Frankfurt, Prof. Dr. Peter Fischer of Institut f\"{u}r Technische Informatik der Universit\"{a}t Heidelberg, Prof. Dr. Peter Senger, CBM Spokeperson and Dr. Subhasis Chattopadhyay of VECC, India for their support in course of this work.


\end{document}